\documentclass{article}
\usepackage{arxiv}

\usepackage{subfigure}
\usepackage[utf8]{inputenc}
\usepackage{tabularx,colortbl}
\usepackage{multirow}
\usepackage{siunitx}
\usepackage{booktabs}
\usepackage{rotating}
              
\usepackage{multicol}

\usepackage{float}
\usepackage{graphicx}
\usepackage{subfiles}
\usepackage{hyperref}
\usepackage{appendix}
\hypersetup{
    colorlinks=true,
    linkcolor=blue,
    filecolor=magenta,      
    urlcolor=cyan,
    citecolor=cyan
}

\usepackage{fancyhdr}
\fancyheadoffset{0pt}
\title{MetaAudio: A Few-Shot Audio Classification Benchmark
\thanks{\textit{Supported by EPSRC, UDRC \& Thales UK. Version 1 (Content as sumbitted to ICANN 2022)}} 
}

\author{
  Calum Heggan\\
  University Of Edinburgh \\
  Edinburgh\\
  \texttt{{s1529508}@sms.ed.ac.uk} \\
   \AND
   Sam Budgett \\
   Thales UK \\
   \texttt{Samuel.BUDGETT@uk.thalesgroup.com} \\
   \And
   Timothy Hospedales \\
   University Of Edinburgh \\
   \texttt{t.hospedales@sms.ed.ac.uk} \\
   \And
   Mehrdad Yaghoobi \\
   University Of Edinburgh\\
   \texttt{m.yaghoobi-vaighan@sms.ed.ac.uk} \\
}

\begin{document}

\maketitle


\begin{abstract}
    \noindent
    Currently available benchmarks for few-shot learning (machine learning with few training examples) are limited in the domains they cover, primarily focusing on image classification. This work aims to alleviate this reliance on image-based benchmarks by offering the first comprehensive, public and fully reproducible audio based alternative, covering a variety of sound domains and experimental settings. We compare the few-shot classification performance of a variety of techniques on seven audio datasets (spanning  environmental sounds to human-speech). Extending this, we carry out in-depth analyses of joint training (where all datasets are used during training) and cross-dataset adaptation protocols, establishing the possibility of a generalised audio few-shot classification algorithm.  Our experimentation shows gradient-based meta-learning methods such as MAML and Meta-Curvature consistently outperform both metric and baseline methods. We also demonstrate that the joint training routine helps overall generalisation for the environmental sound databases included, as well as being a somewhat-effective method of tackling the cross-dataset/domain setting. 

\end{abstract}


\keywords{Few-shot classification \and Meta-learning \and Benchmark \and Audio \and Acoustics}


\section{Introduction}
\label{section:intro}

\noindent
To date, the majority of the breakthroughs seen in machine learning have been in domains or settings where there was an abundance of labelled data, either real or simulated, for example in \cite{alpha_fold}. In contrast, the human capability to  recognise and discriminate between classes of sensory inputs with few examples, e.g. in visual or acoustic settings, remains unmatched. The development of techniques that can perform such Few-shot Learning (FSL) tasks has seen significant interest within modern machine-learning literature, with particular focus on applying meta-learning (learning to learn) \cite{hospedales20201metaSurveyPAMI}. These approaches aim to address the setting where classes are rare or labelled data is hard to produce or gather.

Most work on these types of algorithms focuses on the image domain, with other data modalities and problem settings largely underrepresented. This potentially biases meta-learning algorithmic development towards images, hampering the development of general purpose meta-learners, as well as impeding development of high performance few-shot learning for other types of data or task. Compounding this, the most commonly evaluated datasets, e.g. miniImageNet \cite{matching_networks}, as well as current benchmarks suffer from lack of real-world challenge.

Acoustic classification and event detection have been well studied in conventional fully supervised machine learning \cite{ast,esc}, with many public datasets having a common evaluation protocol that is adhered to by the community, allowing for standardisation and fair comparison. This has however not extended to the few-shot equivalent, where the majority of the works that do exist make little attempt at preserving reproducibility, typically with respect to dataset management and lack of public source code \cite{ml_FSL_SER,fs_aed}. This absence of standardisation poses significant issues when looking to compare novel and existing methods alike.

In this work, we look to alleviate this gap by contributing the following: 1) Experimental evaluation of some of the most popular few-shot classifiers on a variety of audio datasets, spanning multiple sub-settings from environmental sounds to speech. 2) A fully reproducible few-shot audio classification benchmark with at least one published evaluation split per dataset along with custom data loading allowing for quick plug and play testing in future works. 3) A generalised prescription for dealing with variable length audio datasets in a few-shot setting. 4) Finally, in-depth analyses and evaluation of the joint training and cross-dataset/domain settings. We include all of our code 
at \url{https://github.com/CHeggan/MetaAudio-A-Few-Shot-Audio-Classification-Benchmark}.


\section{Few-Shot Classification}
\label{section:few_shot_classification}


\subsection{Formulation}
Generally, few-shot learning involves training, validating and testing a model on pairwise disjoint sets of classes (e.g. classes of human non-speech sounds such as sneezing and coughing), $\mathcal{C}_{train}\notin \mathcal{C}_{val} \notin \mathcal{C}_{test}$. These sets of classes are analogous to the training, validation and test data splits found in traditional machine learning, where splits have non-overlapping samples. In few-shot learning, splits are defined with the addition of non-overlapping classes. 
The goal of a few-shot classifier is to generalise to a set of $\mathnormal{N}$ novel classes, given only a few-labelled examples from each class. These episodes (also referred to as tasks throughout) contain a support set $\mathcal{S}$ which is used for training and a query set $\mathcal{Q}$ where recognition performance is evaluated. 
Meta-learning can either be trained with episodic training \cite{matching_networks}, where individual few-shot tasks are drawn from $\mathcal{C}_{train}$, or non-episodic training \cite{ss}, where a classifier is trained on all classes contained in $\mathcal{C}_{train}$ in order to learn an embedding in the second to last network layer.

\subsection{Meta-Learners \& Other Approaches}
Due to the number of meta-learning algorithms created and distributed in recent literature, discussed more in Section \ref{section:related_work}, we restrict our attention to a  representative few. Specifically these are; Prototypical Networks \cite{proto_nets}, Model-Agnostic Meta-Learning \cite{maml}, Meta-Curvature \cite{MC}, SimpleShot \cite{ss} and Meta-Baseline \cite{mb}. This selection covers both metric and gradient-based meta-learning, as well as extensions to simpler baseline methods. We leave the specific details of the algorithms to the original papers and instead offer a very high level overview. 

\textbf{Prototypical Networks}\quad The most seminal of the metric learning family of meta-learners, ProtoNets \cite{proto_nets} work by calculating class prototypes as the centroid of the embedded support set during learning, followed by applying a nearest-centroid procedure for classifying queries.

\textbf{MAML \& Meta-Curvature}\quad These gradient-based approaches \cite{maml,MC} aim to learn a transferable initialisation for any model such that it can quickly adapt to a new task $\tau$ with only a few steps of gradient descent. At training time, the meta-objective is defined as query set performance after a few steps of gradient descent on the $\mathnormal{K}$ support samples from the model's initial parameters. Meta-curvature expands on MAML by also learning a transform of the inner-optimisation gradients so as to achieve better generalisation on new tasks. In this work, we experiment and report results with first order variants of these algorithms, as in initial experimentation comparing both variants we observed negligible or negative effects to performance.

\textbf{SimpleShot \& Meta-Baseline}\quad SimpleShot \cite{ss} and Meta-Baseline \cite{mb} are simple baseline methods that aim at lowering computational cost while still achieving strong performance. Both methods train in a conventional way, outputting logits directly from a linear layer of size $|C_{train}|$, and validate/test using nearest centroid classification. To distinguish themselves, SimpleShot applies additional data normalisations at test time, while Meta-Baseline performs episodic fine-tuning, similar to ProtoNets but with cosine distance and logit scaling.

\section{Related Work}
\label{section:related_work}

\textbf{Few-Shot Classification}\quad We review only a small subset of available meta-learners and point the reader to \cite{hospedales20201metaSurveyPAMI} for a more detailed review. MAML \cite{maml}, Meta-SGD \cite{meta_sgd} and Meta-Curvature \cite{MC} are representative gradient-based meta-learning (GBML) schemes, designed around the idea of fast adaptation to new learning tasks using additional gradient descent steps. The  prototypical networks \cite{proto_nets} that we evaluate here are metric learners, which aim to learn a strong feature embedding space such that support and queries can be compared using nearest neighbour. Other members of this family are Matching Networks \cite{matching_networks}. All of these algorithms have been primarily evaluated in the image domain and their performance in other domains, audio included, is largely unknown.

\textbf{Few-Shot Acoustics}\quad Currently only a handful of studies exist that look at either few-shot audio classification or event detection. Of these, two are set in event detection \cite{fs_aed,ml_FSL_SER} (classification of parts of an audio clip in time) with the other two focused on classification \cite{transient_classification,fs_classification_study} (classification of an entire audio clip), the focus of this work. Comparing these works, we see a variety of approaches taken toward dataset processing, split formulation and reproducibility. These variations, most importantly the dataset and its associated split used,  make comparisons and ranking of the works impossible. Among these, \cite{transient_classification} is distinct in that it provides both a fully reproducible code base and the dataset class-wise splits used for its experiments. Their main contribution is fitting common metric based learners with an attention similarity module, attached to its purely convolutional backbone. This work is currently state-of-the-art for both the ESC-50 \cite{esc} dataset and its proprietary noise injected variant `noiseESC-50'. As discussed more in Section \ref{section:meta_audio_setup}, we use this work as a basis for some of our experiments. 

\textbf{Benchmarks}\quad Most relevant to our work are other few-shot and meta-learning benchmarks. Included in this are works such as Meta-Dataset \cite{meta_dataset} (an aggregation of 10 few-shot image based datasets) and MetaCC \cite{channel_coding} (a modifiable set of channel coding tasks). Of the benchmarks currently available for few-shot classifier evaluation, none deal with acoustic classification. This is the primary area that this work aims to fill. Meta-Dataset is of particular relevance to this work as we aim to mimic both the depth and reproducibility achieved by the benchmark. Specifically, both the within and cross dataset evaluations as well as the public leaderboard are components which we find to be useful. 


\section{MetaAudio Setup}
\label{section:meta_audio_setup}


\subsection{Setting \& Data}
As MetaAudio aims to be a diverse and reproducible benchmark, it covers a variety of experimental settings, algorithms and datasets. Throughout, we mainly consider 5-way 1-shot classification, with some additional analysis of the impact of k-shots and N-ways at test time. We experiment with 7 total datasets, 5 of which are primary datasets which we split  for use in training and evaluation, and 2 held-out sets we use exclusively for testing. Among them, 3 have fixed-length and 4 have variable length clips. Additional details about the datasets, including size and settings, can be found in Table \ref{table:dataset}. Due to the highly variable sample size of the original dataset and the issues that it presents with reproducibility, we primarily experiment with a pruned version of BirdClef 2020, where samples longer than 180s are removed along with classes with fewer than 50 samples. 

\textbf{Splits \& Labels}\quad For every experiment setup, we apply a 7/1/2 train-validation-testing split ratio over all the classes belonging to an individual dataset. These ratios are chosen to be in line with the majority of machine learning and few-shot works. Any conventional sample based train/val/test splits are ignored, and the class splits are applied to all available data. Outside \cite{transient_classification}, from which we can obtain a reproducible split of ESC-50, we have no works with prior dataset splits to follow, and so we define our own. Most simply we assign random splits based on the available classes for a given set. However, we also define within-dataset domain-stratification and shift splits for sets that have additional internal structure and/or accompanying meta-data. Extensive experimentation with these more specific splits is not carried out in this work, however we include them in our code  \href{https://github.com/CHeggan/MetaAudio-A-Few-Shot-Audio-Classification-Benchmark}{repository}. Labels for the datasets vary quite significantly with some having time strong (temporally localised) labels like BirdClef2020 with others having only weak (clip-level) labels. In the interest of consistency, for this work we drop the available strong labels for the datasets that have them and operate exclusively with weak labels. The tradeoff of this approach is that for datasets that have access to strong labels, we expect additional label noise to be present during training, possibly hurting final generalisation performance. 

\textbf{Pre-Processing}\quad Pre-processing is kept minimal, with only the conversion of raw audio samples into spectrograms and some normalisation factors applied during loading. We consider a fixed sample rate and spectrogram parameters over all datasets and contained samples. For normalisation, three techniques were considered; per sample, channel wise and global. Following initial experimentation, global, which uses average statistics across all examples, was used in all experiments due to performance and simplicity.
\setlength{\tabcolsep}{6pt} 
\renewcommand{\arraystretch}{1.1} 
\begin{table}[t]
  \caption{High level details of all datasets considered in MetaAudio}
  \centering
   \resizebox{\textwidth}{!}{%
  \begin{tabular}{ccccccc}
    \toprule
    Name & Setting & $N^o$ Classes & $N^o$ Samples & Format & Sample Length & Use\\
    \midrule
    ESC-50        & Environmental         & 50   & 2,000   & Fixed    & 5s & Meta-train/test\\
    NSynth        & Instrumentation       & 1006 & 305,978 & Fixed    & 4s & Meta-train/test\\
    FDSKaggle18   & Mixed                 & 41   & 11,073  & Variable & 0.3s - 30s & Meta-train/test\\
    VoxCeleb1     & Voice                 & 1251 & 153,516 & Variable & 3s - 180s & Meta-train/test\\
    BirdCLEF 2020 & Bird Song             & 960  & 72,305  & Variable & 3s - 30m & Meta-train/test\\
    BirdCLEF 2020 (Pruned) & Bird Song    & 715  & 63,364  & Variable & 3s - 180s & Meta-train/test\\
    \midrule
    Watkins Marine Mammal Sounds        & Marine Mammals         & 32   & 1698   & Variable    & 0.1 - 150s & Meta-test\\
    SpeechCommandsV2        & Spoken Word       & 35 & 105,829 & Fixed    & 1s & Meta-test\\
    \bottomrule
  \end{tabular}}
  \label{table:dataset}
\end{table}
 
\subsection{Sampling Strategies}
\label{section:sampling}
Throughout MetaAudio, we utilise a variety of sampling strategies for experimentation. The basis of these is our fixed length approach. The steps used for this can be summarised into: 1) Sample a set of N-way classes $\mathcal{C}\mathnormal{_N}$ from the necessary split of dataset $\mathcal{D}$ and 2) For each class in $\mathcal{C}\mathnormal{_N}$, sample both support and query examples, for support the number will be k-shot. 

We extend this fixed length strategy in order to build a method for dealing with variable length sets. Due to how varied sample length is within some of the considered datasets (Table~\ref{table:dataset}), we convert to fixed length representation to avoid the need for specific neural architectures to support variable length inputs and reduce computational requirements. Specifically, we choose to split our variable length samples into $\mathnormal{L}$ length sub-clips, all sharing the same label. This along with the later conversion of the sub-clips to individual spectrograms is done entirely offline, a decision made to avoid bottlenecking during training. 

Combining a variety of datasets in a joint training and/or evaluation routine has been advocated for in the image space \cite{meta_dataset} to evaluate general purpose representation learning, and potentially improve performance through cross-dataset knowledge sharing. We mimic this and expand upon it for the considered acoustic datasets and task. Sampling tasks from the available datasets in this setting can be done in a few distinct ways. We consider this to be an additional area of investigation and compare two variants of task sampling. In \textbf{Free Dataset Sampling} the $\mathnormal{N}$ classes in an episode can be drawn from multiple source datasets. In \textbf{Within Dataset Sampling} each episode first randomly chooses a dataset, and then draws  $\mathnormal{N}$ random classes within that dataset.

During these sampling strategies, we largely ignore the class sample imbalance seen in the majority of the datasets we experiment with, we do this for a few reasons. The first of these is that recent works, such as \cite{class_imbalance}, suggest it is less detrimental in meta-learning than in conventional learning. The second is that, these imbalances allow algorithms to differentiate themselves with respect to how they handle the more difficult setting. One area in which we do experiment with alleviating the effect of this imbalance is in the re-weighting of the loss functions used in the conventional learning parts of the Meta-baseline and SimpleShot algorithms. To create this dataset custom loss, we employ inverse-frequency class weighting, where the class-wise contribution to the loss function is the inverse of the number of samples present in that class.

\section{Experiments}
\label{section:experiments}

\subsection{Settings}
Our experimental design follows prior few-shot works, where after one end-to-end training and evaluation procedure, average classification accuracies are reported along with their 95$\%$ confidence intervals using 10,000 tasks drawn from the test set. For all experiments we use Adam with a non-adaptive learning rate. Due to the limited tuning performed, we expect it to be fairly easy to obtain a specific result marginally better than those presented, however it is important to note that this does not undermine the comparison and experimental settings investigated in this work.

Motivated by the performance gap between the commonly used CNNs and other neural architectures currently present in conventional acoustic learning \cite{ast,fsd_50k}, we briefly investigated the role of the base neural architecture in the few-shot acoustic setting. Due to space restriction, we do not report details here, however our best performing model using MAML and ProtoNets on ESC-50 was a lightweight hybrid CRNN. Due to its relatively low computational cost compared to larger models, we opt for this architecture throughout. Specifically, the CRNN contains a 4-block convolutional backbone (1-64-64-64) with an attached 1-layer non-bidirectional RNN containing 64 hidden units. The number of outputs in the final linear layer is either of size N-way or, in the case of metric learning and baseline methods, 64. 

In the majority of the results presented for variable length datasets, the value of $\mathnormal{L}$ is set to 5 seconds. We chose this value based on preliminary experiments (not included due to space limitations) where, for Kaggle18, $\mathnormal{L} = 5s$ performed best when compared against 1 and 10-seconds. 
Similarly, for the Watkins Mammal Sound Database, $\mathnormal{L}=5$ closely resembles the expected value of the dataset's sample length distribution.
Setting a common value of $\mathnormal{L}$ also allows us to more comfortably facilitate joint training and cross-dataset evaluation without the need for massive padding. 

\subsection{Within-Dataset Evaluation}\label{sec:within}
We first benchmark the algorithms and datasets using a within-dataset protocol: Training and evaluating on datasets independently and further evaluating using the held out testing classes. From Table~\ref{table:headline_results}(a), we first remark that on ESC-50 our ProtoNet with a CRNN backbone performs at least as well as the Protonet-CNN considered in \cite{transient_classification} with the same split. 

Comparing the results Table~\ref{table:headline_results}(a), we make the following further observations: 
(i) Out of the two fixed length sets, ESC-50 appears to be the harder problem, with  much lower accuracy than NSynth. This is somewhat expected given the very clean NSynth data compared to the noisier ESC-50 data. (ii) The variable length datasets appear to provide a harder setting in general, with significantly lower performance than the fixed-length sets. (iii) Comparing the meta-learners, we see that GBML methods generally perform better, with Meta-Curvature taking first place in 4 out of 5 cases and best average rank; and MAML taking first place on Kaggle18 and second best rank overall. In comparison, our metric and baseline algorithms underperform in accuracy despite their better speed at inference time. We propose that this is due to the GBML methods' adaption mechanism, updating feature representation at each meta-test episode, making them particularly useful for tasks with high inter-class/episode variance. Meanwhile the others must rely on a fixed feature extractor that cannot adapt to each unique episode. Overall the fact that the GMBL methods outperform SimpleShot, in reversal of the widely remarked upon results for miniImageNet in \cite{ss}, shows the value of an audio-based benchmark as a complement to popular image based benchmarks in drawing conclusions about general purpose and domain-specific meta-learner fitness. (iv) Finally, we observe that Meta-Baseline was the most competitive non-adaptive approach. This confirms that episodic meta-learning provides benefit over the conventional supervised representation learning in SimpleShot \cite{ss}. 

\setlength{\tabcolsep}{10pt} 
\renewcommand{\arraystretch}{1.2} 
\begin{table}[h!]
    \caption{Main Meta Audio benchmark 5-way 1-shot classification results. Table (a) contains the within-dataset results, where models are trained for each dataset individually and then evaluated with that dataset's test split. Tables (b) and (c) contain results from the joint training scenario, where we train meta-learners over all datasets simultaneously and then evaluate on individual test splits. They differ in that in (b) we only allow training tasks to be sampled using classes from one of the datasets per episode, whereas in (c) we allow cross-dataset task creation. In (b) and (c) the bottom group of `cross' datasets are held out from training and used only for testing.}
    \centering
    \resizebox{\textwidth}{!}{%
        \begin{tabular}{cccccccc}
        & & & & & & & \\
        \multicolumn{8}{ c }{\textbf{\large a) Baseline Within Dataset Results}}\\
        
        \toprule
       & \multicolumn{2}{ c }{\textbf{Dataset}}  & \textbf{FO-MAML} & \textbf{FO-Meta-Curvature} & \textbf{ProtoNets} & \textbf{SimpleShot CL2N} & \textbf{Meta-Baseline}\\ 
        
        \midrule
        
    	& \multicolumn{2}{ c }{ESC-50}  		& 74.66 $ \pm $ 0.42 & \textbf{76.17 $ \pm $ 0.41}   & 68.83 $ \pm $ 0.38 & 68.82 $ \pm $ 0.39 & 71.72 $ \pm $ 0.38 \\ 
       & \multicolumn{2}{ c }{NSynth} 		& 93.85 $ \pm $ 0.24 & \textbf{96.47 $ \pm $ 0.19} & 95.23 $ \pm $ 0.19 & 90.04 $ \pm $ 0.27 & 90.74 $ \pm $ 0.25 \\ 
        & \multicolumn{2}{ c }{Kaggle18}         	& \textbf{43.45 $ \pm $ 0.46} & 43.18 $ \pm $ 0.45 & 39.44 $ \pm $ 0.44 & 42.03 $ \pm $ 0.42 & 40.27 $ \pm $ 0.44 \\ 
        & \multicolumn{2}{ c }{VoxCeleb1}         & 60.89 $ \pm $ 0.45 & \textbf{63.85 $ \pm $ 0.44} & 59.64 $ \pm $ 0.44 & 48.50 $ \pm $ 0.42 & 55.54 $ \pm $ 0.42 \\
       & \multicolumn{2}{ c }{BirdClef (Pruned)} & 56.26 $ \pm $ 0.45 & \textbf{61.34 $ \pm $ 0.46}  & 56.11 $ \pm $ 0.46 & 57.66 $ \pm $ 0.43 & 57.28 $ \pm $ 0.41 \\
        
        \midrule
        
        & \multicolumn{2}{ c }{Avg Algorithm Rank} & 2.4 & 1.2 & 3.8 & 4.0 & 3.6 \\
            
         \bottomrule
        
        & & & & & & & \\
        
        \multicolumn{8}{ c }{\textbf{\large b) Joint Training (Within Dataset Sampling)}}\\
        

         \midrule
         & \multicolumn{2}{ c }{ESC-50}  	& 68.68 $ \pm $ 0.45 & \textbf{72.43 $ \pm $ 0.44} & 61.49 $ \pm $ 0.41 & 59.31 $ \pm $ 0.40  & 62.79 $ \pm $ 0.40\\ 
         & \multicolumn{2}{ c }{NSynth}     &  81.54 $ \pm $ 0.39  & 82.22 $ \pm $ 0.38 & 78.63 $ \pm $ 0.36 & \textbf{89.66 $ \pm $ 0.41} & 85.17 $ \pm $ 0.31\\ 
         & \multicolumn{2}{ c }{Kaggle18}   & 39.51 $ \pm $ 0.44 & \textbf{41.22 $ \pm $ 0.45} & 36.22 $ \pm $ 0.40 & 37.80 $ \pm $ 0.40 & 34.04 $ \pm $ 0.40\\ 
         & \multicolumn{2}{ c }{VoxCeleb1}  & \textbf{51.41 $ \pm $ 0.43} & 51.37 $ \pm $ 0.44 & 50.74 $ \pm $ 0.41 & 40.14 $ \pm $ 0.41  & 39.18 $ \pm $0.39\\
         \multirow{-5}{*}{\rotatebox[origin=c]{90}{\textbf{Trained}}} 
            & \multicolumn{2}{ c }{BirdClef (Pruned)}     & \textbf{47.69 $ \pm $ 0.45} & 47.39 $ \pm $ 0.46   & 46.49 $ \pm $ 0.43 & 35.69 $ \pm $ 0.40 &  37.40 $ \pm $ 0.40 \\
            
         \midrule
        
         & \multicolumn{2}{ c }{Watkins}     & 57.75 $ \pm $ 0.47 & \textbf{57.76 $ \pm $ 0.47} & 49.16 $ \pm $ 0.43 & 52.73 $ \pm $ 0.43 & 52.09 $ \pm $ 0.43\\ 
         \multirow{-2}{*}{\rotatebox[origin=c]{90}{\textbf{Cross}}}& \multicolumn{2}{ c }{SpeechCommands V1}     & 25.09 $ \pm $ 0.40 & \textbf{26.33 $ \pm $ 0.41} & 24.31 $ \pm $ 0.36 & 24.99 $ \pm $ 0.35 & 24.18 $ \pm $ 0.36 \\
        
         \midrule
         
        & \multicolumn{2}{ c }{Avg Algorithm Rank} & 2.0 & 1.6 & 4.0 & 3.4 & 4.0 \\
        \bottomrule
      
        & & & & & & & \\
    \multicolumn{8}{ c }{\textbf{\large c) Joint Training (Free Dataset Sampling)}}\\
        
        \toprule
        
    	& \multicolumn{2}{ c }{ESC-50}  	& \textbf{76.24 $ \pm $ 0.42} & 75.72 $ \pm $ 0.42 & 68.63 $ \pm $ 0.39 & 59.04 $ \pm $ 0.41 & 61.53 $ \pm $ 0.40\\ 
        & \multicolumn{2}{ c }{NSynth}      & 77.71 $ \pm $ 0.41 & 83.51 $ \pm $ 0.37 & 79.06 $ \pm $ 0.36 & \textbf{90.02 $ \pm $ 0.27} & 85.04 $ \pm $ 0.31 \\ 
        & \multicolumn{2}{ c }{Kaggle18}    & 44.85 $ \pm $ 0.45 & \textbf{45.46 $ \pm $ 0.45} & 41.76  $ \pm $ 0.41 & 38.12 $ \pm $ 0.40 & 35.90 $ \pm $ 0.38\\ 
        & \multicolumn{2}{ c }{VoxCeleb1}   & 39.52 $ \pm $ 0.42 & 39.83 $ \pm $ 0.43 & 40.74 $ \pm $ 0.39 & \textbf{42.66 $ \pm $ 0.41} & 36.63 $ \pm $ 0.38 \\
        
        \multirow{-5}{*}{\rotatebox[origin=c]{90}{\textbf{Trained}}} 
            & \multicolumn{2}{ c }{BirdClef (Pruned)}     & \textbf{46.76 $ \pm $ 0.45} & 46.41 $ \pm $  0.46  & 44.70 $ \pm $ 0.42 & 37.96 $ \pm $ 0.40 & 32.29 $ \pm $ 0.38 \\
            
        \midrule
        
        & \multicolumn{2}{ c }{Watkins}     & \textbf{60.27 $ \pm $ 0.47} & 58.19 $ \pm $ 0.47 & 48.56 $ \pm $ 0.42 & 54.34 $ \pm $ 0.43 & 53.23 $ \pm $ 0.43\\ 
        
        \multirow{-2}{*}{\rotatebox[origin=c]{90}{\textbf{Cross}}}& \multicolumn{2}{ c }{SpeechCommands V1}    & \textbf{27.29 $ \pm $ 0.42}  & 26.56 $ \pm $ 0.42 & 24.30 $ \pm $ 0.35 & 24.74 $ \pm $ 0.35 & 23.88 $ \pm $ 0.35 \\
        
        \midrule
        & \multicolumn{2}{ c }{Avg Algorithm Rank} & 2.1 & 2.1 & 3.4 & 3.0 & 4.3 \\
        
        \bottomrule
        
    \end{tabular}}
    \label{table:headline_results}
\end{table}

\subsection{Joint Training \& Cross Dataset} In this section we extend our evaluation to joint training, where a single model is learned on the combined  training splits of all source datasets (rather than a per-dataset model as in Section~\ref{sec:within}), and then evaluated on the testing split of each dataset in turn. Furthermore, we now also test on two held-out datasets that were not included during training (Table~\ref{table:dataset}), to evaluate cross-dataset few-shot learning performance. We report results for both within-dataset and free-dataset episode sampling (as discussed in Section~\ref{section:sampling}) in Table \ref{table:headline_results}(b) and Table \ref{table:headline_results}(c) respectively. 

First, we compare the joint training regime results \ref{table:headline_results}(b,c) against the within-dataset evaluation in Table~\ref{table:headline_results}(a). For both ESC-50 and Kaggle18 we obtain new SOTA results with MAML and Meta-Curvature respectively, both from the free dataset sampling routine. For all other datasets, we see a degradation of performance compared to within dataset training in Table~\ref{table:headline_results}(a). This difference varies in magnitude between datasets and sampling routines. The mixed results here mirror those observed in \cite{meta_dataset} and reflect the tradeoff between two forces: (1) a positive effect of generally increasing the amount of training data available compared to the within-dataset condition, and (2) a negative effect due to the increased difficulty of learning a single model capable of simultaneous high performance on diverse data domains \cite{rebuff2017mdl}. This shows that MetaAudio complements \cite{meta_dataset} in providing a challenging benchmark to test  future meta-learners' ability to fit diverse audio types, as well as enabling few-shot recognition of new categories.

Moving to the other question of interest, we contrast how the joint training episode sampling routines compare. For our main datasets, we observe 3/5 of the top results were obtained using the free sampling method, with the 2 outliers belonging to VoxCeleb and BirdClef - evidence that their tasks require significantly different and specific model parametrisation, as the within dataset task sampling would allow more opportunity to learn these more specialised features. 

For the held-out cross-dataset tasks (Watkins, SpechCommands), we also see the strongest performance coming from the free sampling routine, where it outperforms its within dataset counterpart by $\sim$2\% in both held-out sets. As for the absolute performances obtained on the held-out sets, we see that our joint training transfers somewhat-effectively, with the model in one case attaining a respectable 50-60\% and another obtaining accuracies only 5\% above random. 

Finally, comparing meta-learners, we again see GBML approaches Meta-Curvature and MAML performing best overall. However, for this joint training condition, SimpleShot improves to take third place overall by average rank.

\subsection{External Data \& Pretraining} A full meta-learning pipeline for a specific dataset can be expensive. Recent studies in the few-shot and self-supervision communities have debated whether off-the-shelf models pre-trained on large external datasets may provide a better approach to few-shot than meta-learning \cite{ss}. Transferring a well-trained representation and training a simple classifier for each task could also be cheaper due to amortizing the cost of large-scale pre-training over multiple downstream tasks. 

To this end, we also evaluate our MetaAudio benchmark using pre-trained feature models trained on the large scale ImageNet \cite{imagenet} and AudioSet \cite{audioset} datasets. Specifically, we use the SOTA Audio Spectrogram Transformers(ASTs) from \cite{ast}. We experiment with two model variants, the ImageNet only and the ImageNet + AudioSet trained `base384' transformers provided by the authors, including their suggested AudioSet sample normalisation. We apply both nearest-centroid and SVM linear classification on our output features. 

\setlength{\tabcolsep}{10pt} 
\renewcommand{\arraystretch}{1} 
\setlength{\tabcolsep}{10pt} 
\renewcommand{\arraystretch}{1} 
\begin{table}[h!]
    \caption{Meta Audio benchmark using a variety of pre-trained spectrogram transformers from \cite{ast}. Models are trained on ImageNet \cite{imagenet} and AudioSet \cite{audioset}. Results show 5-way 1-shot performance using simple classifiers on fixed features. We compare these to the results for SimpleShot using dataset specific training and evaluation. }
    \noindent
    \centering
    \resizebox{\textwidth}{!}{%
        \begin{tabular}{cccccc|c}\toprule
	
	\multicolumn{2}{ c }{} & \multicolumn{2}{ c }{\textbf{AST ImageNet}}  & \multicolumn{2}{ c }{\textbf{AST ImageNet \& AudioSet}}  & \\	

        \multicolumn{2}{ c }{\textbf{Dataset}}  & \textbf{SVM} & \textbf{SimpleShot (CL2N)} & \textbf{SVM} & \textbf{SimpleShot (CL2N)} & \textbf{SimpleShot (CL2N) from Table \ref{table:headline_results}} a)\\ 
        
        \midrule
        
    	\multicolumn{2}{ c }{ESC-50}  		    & 61.12 $ \pm $ 0.41 & 60.41 $ \pm $ 0.41 & 61.61 $ \pm $ 0.41 & 64.48 $ \pm $ 0.41 &  \textbf{68.82 $ \pm $ 0.39} \\  
        \multicolumn{2}{ c }{NSynth} 		    & 64.26 $ \pm $ 0.41 & 66.68 $ \pm $ 0.41 & 62.62 $ \pm $ 0.42 & 63.78 $ \pm $ 0.42 & \textbf{90.04 $ \pm $ 0.27}\\  
        \multicolumn{2}{ c }{Kaggle18}         	& 34.01 $ \pm $ 0.40 & 33.52 $ \pm $ 0.39 & 38.38 $ \pm $ 0.41 & 38.76 $ \pm $ 0.41 & \textbf{42.03 $ \pm $ 0.42}\\  
        \multicolumn{2}{ c }{VoxCeleb1}         & 27.26 $ \pm $ 0.36 & 28.09 $ \pm $ 0.37 & 27.45 $ \pm $ 0.36 & 28.79 $ \pm $ 0.38 & \textbf{48.50 $ \pm $ 0.42} \\
        \multicolumn{2}{ c }{BirdClef (Pruned)} & 30.84 $ \pm $ 0.37 & 33.04 $ \pm $ 0.41 & 33.17 $ \pm $ 0.38 & 36.41 $ \pm $ 0.42  & \textbf{57.66 $ \pm $ 0.43}\\
        \midrule 
        \multicolumn{2}{ c }{Avg Rank} & 4.2 & 3.8 & 3.6 & 2.4 & 1.0\\
        \midrule 
        \multicolumn{2}{ c }{Watkins}           & \textbf{55.91 $ \pm $ 0.42} & 55.40 $ \pm $ 0.42 & 51.46 $ \pm $ 0.42 & 51.81 $ \pm $ 0.42  & N/A\\
        \multicolumn{2}{ c }{SpeechCommands V1} & 26.24 $ \pm $ 0.36 & 26.46 $ \pm $ 0.37 & \textbf{30.69 $ \pm $ 0.38} & 30.24 $ \pm $ 0.38  & N/A\\
        \midrule
        \multicolumn{2}{ c }{Avg Rank} & 2.5 & 2.5 & 2.5 & 2.5 & N/A\\
        
        \bottomrule
    \end{tabular}}
    \label{table:ast_pretrain}
\end{table}
The results in Table~\ref{table:ast_pretrain} show that the features pre-trained on AudioSet and ImageNet unsurprisingly outperforms those pre-trained on ImageNet alone. However, the small size of this margin is perhaps surprising, showing that image-derived features provide most of the information needed to interpret spectrograms. 

Comparing these results to in-domain training in Table~\ref{table:headline_results}, we see that performance has dropped substantially, with the potential exception of ESC-50 and Kaggle18. In their best cases, NSynth, VoxCeleb and BirdClef all take drops in performance of $\sim$20\% due to dataset shift between general purpose pre-training and our specific tasks, such as musical instruments, speech or bird song recognition. While the performance hit due to domain-shift is expected, these results are surprising as AudioSet is a much larger dataset, and the AST transformer is a much larger architecture than the CRNN used in Table~\ref{table:headline_results}. In image modality, analogous experiments show a clear win simply applying larger pre-training datasets and larger-models combined with simple readouts, compared to conducting within-domain meta-learning \cite{dumoulin2021a}. This confirms the value of Meta-Audio as an important benchmark for assessing meta-learning contributions that cannot easily be replicated by larger architectures and more data. Performance on our held-out sets shows a more mixed set of results, with ImageNet only pre-training favouring Watkins, and ImageNet + AudioSet pre-training setting a new SOTA for SpeechCommands. 

\subsection{N-Way k-Shot Analysis}
\begin{figure}[h!]
    \centering
    \includegraphics[width=0.75\textwidth]{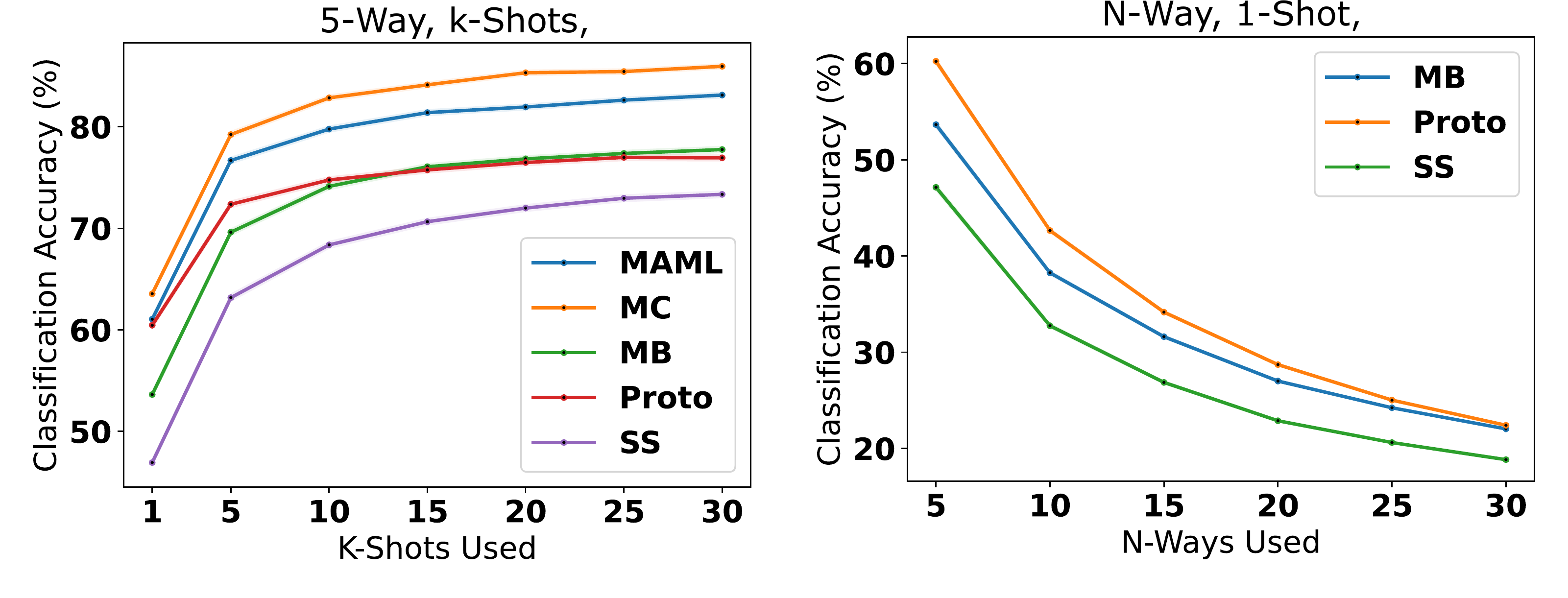}
    \caption{Few-shot learning on VoxCeleb1. Varying test K-shots (left) and N-ways (right).}
    \label{figure:n_way_k_shot}
\end{figure}

Although we only trained and evaluated on the task of 5-way 1-shot classification, we are interested in the effect of larger shots and wider ways on algorithm performance. To bridge this gap, we experiment with these components at test time, using our already trained 5-way 1-shot models. We consider all of our primary datasets and algorithms, covering values of N from 5-30, and k from 1-30. Varying N-ways and k-shots are treated separately and not stacked, a decision made to avoid the compounding computational complexity of the problem. For algorithms which have a fixed size output (i.e. GBML methods) we exclude the varying N-ways. Both ESC-50 and Kaggle18 have only 10 and 7 classes belonging to their test sets respectively, and so analysis further than 10/5-way is impossible. We include a sample of our result plots in Figure \ref{figure:n_way_k_shot}. 
Varying the numbers of shots, we observe a clear trend of GBML methods outperforming baseline and metric learning approaches. This is especially true for large k-shots, where the rise in performance also occurs faster. For both fixed length sets, we see some additional distinction between gradient based methods and the others, where methods without adaptation both stagnate and start to decline in performance after 5-shot. Up to 30-shot, we do not observe this same behaviour in variable length sets, however it is possible that this is simply due to the complexity of the problems. Of the three non-gradient-based methods, which algorithm performs best over k-shots appears to be dataset specific, with each outperforming the others in at least one set. Although we are more limited in varying the number of ways we test over, we still observe some interesting trends. All of our tested algorithms show a non-linear decay in performance, with results at 30-way still reaching $\sim$20-25\% for our VoxCeleb and BirdClef sets (approx 7$\times$ random). For speed of drop-off, we see a similar story as we saw in increasing k-shot, with all algorithms showing the best performance in at least one set as N-way increases.


\section{Conclusion}
\label{section:conclusion}

In this work, we presented MetaAudio, a new large-scale and diverse few-shot acoustic classification benchmark covering a variety of algorithms, sound domains and experimental settings.

Our experiments showed that gradient-based meta-learners with feature adaptation capability generally performed better than fixed-representation competitors. This was the case across most of our settings, although the latter algorithms benefitted from faster learning speed.

We also evaluated the ability of meta-learners to span few-shot learning tasks drawn from a heterogeneous variety of datasets, and their ability to generalise across distribution shift between training and testing. Surprisingly, we also showed that in-domain meta-learning led to substantially better performance than transfer learning from larger architectures trained on larger external datasets, a result that is noticeably different to that from computer vision. 

Going forward, MetaAudio will provide a substantial complement to influential analogous benchmarks \cite{meta_dataset} in the vision domain. Besides benefiting few-shot learning in audio domain, we believe that MetaAudio will help to drive meta-learning research overall, ensuring that more generally relevant algorithms and insights are developed, without becoming overly-specific to computer vision (a problem of gaining relevance). This should help ensure meta-learning research benefits data efficient learning demand across society more broadly.


\section*{Acknowledgments}
\label{section:conclusion}

This work is supported by the Engineering and Physical
Sciences Research Council of the UK (EPSRC) Grant number
EP/S000631/1 and the UK MOD University Defence Research
Collaboration (UDRC) in Signal Processing, EPSRC iCASE account EP/V519674/1 and Thales UK Ltd. 

\bibliographystyle{unsrt}  
\bibliography{refs.bib}

\end{document}